\documentclass[11pt,tightenlines,eqsecnum,floats,aps,nofootinbib,prd]{revtex4-2}
\usepackage{hyperref}
\usepackage{amsmath,amssymb,amsfonts,amsthm,amscd}
\usepackage{enumerate}
\usepackage{colordvi}
\usepackage{units}
\usepackage{epsfig}
\usepackage{natbib}
\usepackage{enumerate}
\usepackage{afterpage}
\usepackage[utf8]{inputenc}
\def\ba{\begin{eqnarray}}
\def\ea{\end{eqnarray}}
\def\be{\begin{equation}}
\def\ee{\end{equation}}

\newcommand{\pv}[1]{\frac{\partial}{\partial #1}}
\def\Lie{\mathfrak{L}}
\def\L{\mathcal{L}}
\def\H{\mathcal{H}}
\def\D{\mathbf{D}}

\begin{document}

\title{A New Action for Cosmology}
\author{David Sloan}
\email{d.sloan@lancaster.ac.uk}
\affiliation{Department of Physics, Lancaster University, Lancaster UK}

\begin{abstract}
\noindent We present a new action which reproduces the cosmological sector of general relativity in both the Friedmann-Lema\^itre-Robertson-Walker (FLRW) and Bianchi models. This action makes no reference to the scale factor, and is of a frictional type first examined by Herglotz. We demonstrate that the extremization of this action reproduces the usual dynamics of physical observables, and the symplectification of this action is the Einstein-Hilbert action for cosmological models. We end by discussing some of the increased explanatory power produced by considering the reduced physical ontology resulting from eliminating scale.  

\end{abstract}

\maketitle

\section{Introduction}
\label{Sec:Introduction}

Modern cosmology takes as its foundation symmetry reduced models of general relativity, typically treating space as homogeneous (and often isotropic). Such reductions take a system of partial differential equations and produce a set of simple ordinary differential equations in their place. The dynamics of these models is encoded in variables that describe the size of the universe, invoking the use of a fiducial cell when dealing with non-compact manifolds. The evolution of this size, represented by the scale factor, is the basis of most descriptions. 

It is well known that the scale factor alone, despite its apparent central role, carries no physical meaning. It is only relative change in the size which it represents that has any effect upon physical observables. For example the temperature of the cosmic microwave background observed by the Planck satellite is a simple product of its temperature when the universe first became transparent, shortly after the time of recombination, and the ratio of the scale factor today to that at the time of this transparency. The Friedmann equations
\be H^2 + \frac{k}{a^2} = \frac{8\pi G \rho}{3} \quad \quad \dot{H}+H^2 = - \frac{4 \pi G}{3} (\rho + 3P) \ee 
relate the expansion rate $H=\dot{a}/a$ to the energy density, $\rho$, and pressure, $P$, of the matter present. These are insensitive to a change in which $a \rightarrow \lambda a, k \rightarrow \lambda^2 k$. Furthermore, the equations of motion for the matter variables are also insensitive to the overall value of the scale factor. As an example the continuity, or fluid, equation
\be \dot{\rho}+3H(\rho+P) \ee
is indifferent to the scale factor. This turns out to be a general feature of all minimally coupled matter models, and should be considered unsurprising. In fact the property is much more general than this - any coupling to scale can be treated similarly through the introduction of new kinetic terms, as we shall see later. For simplicity of exposition we will restrict ourselves to the case of minimally coupled matter in this paper. The geometry of a space-time is inferred from the behaviour of matter within it, and thus to an observer who inhabits such a universe transformations of the geometry that have no effect on the behaviour of the matter within it will be indistinguishable. In the case of a non-compact spatial slice all measurements of the scale factor are made with respect to a fiducial cell and physics should be indifferent under changes of such a cell.  

Commonly attributed to Leibniz, the Principle of the Identity of Indiscernibles (PII) states that two mathematical entities $O$ and $O'$ which give rise to the same set of observations should be considered to be a single physical system \cite{PII}. To quote Weyl, when considering a map between two such entities ``Only such relations will have objective meaning as are independent of the mapping chosen and therefore remain invariant under deformations of the map." \cite{weyl2009philosophy}. As such we should consider two cosmological models whose only differences are the value of the scale factor (and curvature $k$) at some event to describe the same physical system, and only invariants of this transformation should have physical meaning. Thus in the mathematical representation of cosmology the choice of scale factor constitutes a redundancy. On the surface level, this may seem unproblematic - the redundancy would appear to be no impediment to describing a large set of cosmological solutions, and indeed can be used to aid our intuitive understanding. However there are ontological consequences of this extraneous structure. The counting of models as being distinct through differing scale factors leads to an infinite measure on the space of FLRW cosmologies, and thus notions of typicality or probability become ill defined \cite{Gibbons:1986xk,Measure,Corichi:2010zp,Measure2,Corichi:2013kua}, leading to interesting behaviour on relational spaces\cite{Barbour2014,Barbour2015,FlavioSDbook}. Further, the physical intuition we gain from associating the scale factor with size becomes problematic as we approach a singularity where typically the this becomes zero. This is the root cause of systems becoming non-predictive at the big bang, and it has been recently shown that working in systems without the scale factor these solutions can be continued predictively beyond this point \cite{Through}. Finally, the intuition we gain from introducing the scale factor is tempered by its making cosmological systems conservative. As we shall see, working without it we can understand cosmological dynamics as being frictional, and thus thermodynamically richer in explanatory power than their conservative counterparts. The scale factor is a vestigial structure left over from the geometric description of general relativity. The question then arises as to whether we can reproduce cosmological dynamics from an action that makes no reference to such a structure.

In fact, a complete dynamics of the cosmological system can be derived from an action principle that makes no reference to the scale factor at all. Let us consider the case of matter described by a Lagrangian $\L_m$ which is minimally coupled, in the context of a flat FLRW cosmology. This will be generalized in later sections to include curvature and anisotropies. Then the action principle is to minimize the action, $S(t)$, at some time where:
\be S(t) = S_0 + \int_0^t \left(\frac{3S^2}{2} + 4\pi G \L_m \right) dt' \ee
subject to initial conditions for the variables. In other words the action, $S$, obeys:
\be \dot{S} = \frac{3S^2}{2} + 4 \pi G \L_m (\vec{q},\dot{\vec{q}}) \label{FLRWHerglotzAction} \ee
To reproduce the cosmological dynamics we identify $S$ with minus of the Hubble parameter, $S=-H$. Extremizing this action we find the equations of motion for the matter fields, $\vec{q}$ are given:
\be 
\frac{d}{dt} \left( \frac{\partial \L_m}{\partial \dot{q_i}} \right) - \frac{\partial \L_m}{\partial q_i} +3H \frac{\partial \L_m}{\partial \dot{q_i}} = 0
\ee
which are the same equations of motion as are found from the Einstein-Hilbert action minimally coupled to matter. For example, taking $\L_m$ to describe a scalar field with a potential we would recover the usual Klein-Gordon equation familiar to inflationary cosmology. Further, since $S=-H$, equation (\ref{FLRWHerglotzAction}) is the equation of motion for the Hubble parameter given by the Einstein-Hilbert action. Thus the dynamics of the physically observable quantities (the Hubble parameter and the matter content) are identical to those derived from general relativity. 

There are several things of note about this action. The first, and most striking, is that the evolution of the action is dependent upon the value of the action itself. This stands in stark contrast to the usual Euler-Lagrange formulations of physics, and is a generalization that was first considered by Herglotz. We examine such actions in more detail in section \ref{Sec:Herglotz}. The second thing of note is that equation (\ref{FLRWHerglotzAction}) is the Raychaudhuri equation, and thus our principle is that we minimize the Hubble parameter subject to this. This suggests an extension beyond cosmological systems - the minimization of the extrinsic curvature of a spatial slice subject to the Raychaudhuri equation. A third area of interest is that the equation of motion (\ref{FLRWHerglotzAction}) depends on an odd number of variables - the matter configuration variables, their velocities and the action itself. This does not allow for a symplectic description of the theory as any phase space is even dimensional. As we will show, there exists a symplectification of the theory which turns out to be the usual Einstein-Hilbert action adapted to Robertson-Walker metrics. However when working with the action directly we see that the variables live on a contact manifold. These are described in further detail in section \ref{Sec:Herglotz}. 

This paper is laid out as follows: In the following section (section \ref{Sec:Herglotz}) we describe a general form of action dependent Lagrangians, which we will call `Herglotz Lagrangians'. We will examine the general behaviour of these, their extension through `symplectification' to a system that recovers a conservative description by introducing more degrees of freedom, and give and example in the form of a damped harmonic oscillator. In section \ref{Sec:FLRW} we show how this applies to the action described above, and how this can include curvature terms. The extension to anisotropic cosmologies is given in section \ref{Sec:Bianchi}, which details the action principle for Bianchi models. In section \ref{Sec:AndBackAgain} we show how the dynamical similarity of cosmological systems can be used to reveal our action from the usual Einstein-Hilbert action, and we discuss the implications of our findings is section \ref{Sec:Discussion}.

\section{Herglotz's Principle}
\label{Sec:Herglotz}

In this section we will briefly recap Herglotz's variational principle and some pertinent results relating to the contact manifolds on which dynamics takes place. There are several extant papers describing these in greater detail, so here we will simply cover the necessary material to describe cosmological systems. Throughout we will work with a damped harmonic oscillator as an example of a system that can be described this way and demonstrates the features that emerge from our cosmological system. Herglotz considered a generalization of Lagrangian dynamics in which the evolution of the system could depend on the action itself. This is a necessity in studying some non-conservative systems, as it was proved by Bauer that the normal action principles of Euler and Lagrange cannot have dissipative terms proportional to a velocity. Herglotz examined systems whose evolutions can be expressed as
\be \dot{S}=\L^{H} (q_i,\dot{q_i},S) \label{HerglotzDefinition} \ee
wherein we draw attention to the close relation to Lagrangian theories by noting that the evolution of $S$ is a ``Herglotz Lagrangian" through the $H$.\footnote{Herglotz considered a more general, time dependent case wherein $\L^H$ could also depend on $t$. However we will not require this level of generality and thus stay with the simpler case described here.} Note that unlike the usual Lagrangian, to find $\L^H$ we require not only the tangent bundle over the configuration space, $TM$, but its extension which forms an odd-dimensional contact space \cite{geiges2008introduction,ContactIntro}. We note that if $\L^H$ is independent of $S$ then a trivial integration over time shows that $S$ forms a normal action principle with $\L^H$ its Lagrangian. In this generalized system the equations of motion are derived by extremizing the action $S$ at some time, subject to equation (\ref{HerglotzDefinition}) and initial conditions. To find the equations of motion that arise from this extremization, consider the more general action $A$ where we integrate the Herglotz Lagrangian and enforce equation (\ref{HerglotzDefinition}) through a Lagrange multiplier $\lambda$ which is a general function of time:
\be A(q,\dot{q},S,\dot{S},t) = \int \left( \L^H + \lambda(\L^H - \dot{S}) \right) dt \ee
It is clear that when equation (\ref{HerglotzDefinition}) is satisfied $A=S$ and so the extremization of $A$ will extremize $S$. The extremization of $A$ gives the usual Euler-Lagrange equations for $q$ and $S$ in terms of the Lagrange multiplier:
\be \frac{d}{dt} \left( (1+\lambda) \frac{\partial \L^H}{\partial \dot{q}} \right) = (1+\lambda)\frac{\partial \L^H}{\partial q} \quad \frac{d\lambda}{dt}  = -(1+\lambda) \frac{\partial \L^H}{\partial S} \ee
Combining these using the second to replace $\dot{\lambda}$ in the first, and noting that, since $\lambda$ is a Lagrange multiplier, $1+\lambda$ is generally non-zero, we find the equation of motion:
\be 
\frac{d}{dt} \left(\frac{\partial \L^H}{\partial \dot{q_i}}\right) - \frac{\partial \L^H}{\partial q_i} - \frac{\partial \L^H}{\partial S} \frac{\partial \L^H}{\partial \dot{q_i}} = 0 \label{HerglotzGeneralEOM} 
\ee
These equations reduce to the expected Euler-Lagrange equations when $\L^H$ is independent of $S$. 

Typically a Herglotz Lagrangian is used to describe non-conservative systems by introducing friction. As an example consider the system described by 
\be \L^H = \frac{m \dot{x}^2}{2} - \frac{k x^2}{2} - \frac{\mu S}{m} \label{DampedHarmonicHerglotzLagrangian} \ee
From this we find the equation of motion for a damped harmonic oscillator:
\be m \ddot{x} + \mu \dot{x} + k x = 0 \ee
Thus we see that the additional term we have introduced has caused the system to be non-conservative. In cosmological terms this makes more precise that nature of ``Hubble friction" - the analogy is direct. The conservative nature of typical Lagrangian systems is captured by the existence of a conserved Hamiltonian. It is interesting to note that typically to introduce friction into a Hamiltonian system we would need to have a reservoir or second physical system into which energy can be exported. The conservative nature of these means that any loss from one system would be gained by the second, and thus the overall description would remain conservative. This is a particularly important point which we will describe in more detail when considering the cosmological implications, as the universe is broadly considered to be an isolated system. We see that the equivalent of the Hamiltonian derived from this Herglotz Lagrangian, a contact Hamiltonian, is not conserved. To find the contact Hamiltonian we perform a Legendre transform on the Herglotz Lagrangian. In doing so, we move our description from the tangent bundle to the cotangent bundle in the usual manner for the variables $q,\dot{q} \rightarrow q,p=\frac{\partial \L^H}{\partial \dot{q}}$, but retain $S$ unchanged. Thus the contact Hamiltonian, $\H^c$, is given:
\be \H^c = p\dot{q}- \L^H \ee
We note here that we are using the symplectic structure of $T^*M$ in this construction. Also we will consider here only the case where this transformation is well-defined due to the form of $\L^H$. Generalizations are simple in the case of first and second class constraints, which we will leave to more technical descriptions such as \cite{Bravetti,Bravetti2,Leon}. Thus in our contact Hamiltonian system dynamics takes place on a contact manifold, $\mathcal{C^*} = \mathbb{R} \times T^*M$ on which we can take $S,q,p$ as coordinates. The equations of motion for such systems in these Darboux coordinates are given \cite{ArnoldBook}:
\be 
\dot{q} = \frac{\partial \H^c}{\partial p} \quad \dot{p} = - \frac{\partial \H^c}{\partial q} - p \frac{\partial \H^c}{\partial S} \quad \dot{S} = p \frac{\partial \H^c}{\partial p}-\H^c 
\ee
wherein we note that the equation for $\dot{S}$ reproduces the Herglotz Lagrangian, and to generalize to multiple variables should be summed over all the momenta - $\dot{S}=\sum_i p_i \frac{\partial \H^c}{\partial p_i} - \H^c$. The equations for $\dot{q}$ and $\dot{p}$ reduce to the usual Hamilton's equations when $\H^c$ is independent of $S$. The presence of the second term in the equation for $\dot{p}$ is the manner through which friction is manifest in the system. Together these can be used to show that the contact Hamiltonian is not time independent except when it is zero (or independent of $S$):
\be \dot{\H^c} = -\H^c \frac{\partial \H^c}{\partial S} \ee
and hence we see the non-conservative nature of the system in general in terms of energy loss.  For the damped harmonic oscillator example introduced above we see that this results in 
\be \dot{\H^c} = -\mu \H^c \rightarrow \H^c= E e^{-\frac{\mu t}{m}} \label{DampedHamiltonianEvolution} \ee
for some constant $E$. Cosmological systems are contained within the exception here, however, since the contact Hamiltonian is constrained to be zero.

A second way in which the non-conservative nature of these systems is manifest is through the evolution of a volume form on $\mathcal{C}^*$. For regular Hamiltonian systems, Liouville's theorem states that the symplectic structure, $\omega$ is preserved under time evolution. Hence on $T^*M$ a volume form composed by taking $\omega^{\wedge n}$ wherein $n$ is the number of configuration variables, retains its size under the Hamiltonian flow. However, $\mathcal{C}^*$ is an odd dimensional manifold, and since $\omega$ is a two-form, it is not possible to create a volume form in the same manner. The analogue of these structures is the contact form, expressed in Darboux coordinates as $\eta = -dS + p dq$. This is a one form, and bears a striking resemblance to the symplectic potential, $\theta$, where $\omega = d\theta$. A canonical choice of volume form on $\mathcal{C}^*$ is $\Omega = \eta \wedge d\eta^{\wedge n}$, and the corresponding Liouville-type theorem is that 
\be \dot{\Omega} = -(1+n) \frac{\partial \H^c}{\partial S} \Omega \ee
Hence we see that over time solutions can focus on areas of $\mathcal{C}^*$ - called attractors. The behaviour of these is described at length in \cite{DynSim}. 

One can embed a contact system within a conservative one through symplectification. To do so first let us form the symplectic manifold by extending the contact manifold. The contact form $\eta$ is defined only up to a choice of overall scale, and an equivalent form can thus be produced by taking $\eta'=y \eta$, for $y \in \mathbb{R}$. We can then promote $y$ to a coordinate and form the extended symplectic system $\mathbb{R} \times \mathcal{C}^*$ with symplectic structure $\omega = d(y\eta) = dS \wedge dy + d(yp) \wedge dq$. Hence the new momentum conjugate to $q$ on this manifold is $\pi=yp$. This construction is equivalent to taking the product of the configuration manifold with $\mathbb{R}$ (coordinatized through $y$) and finding the cotangent bundle. Then we form the Hamiltonian on this manifold by taking $\H=y\H^c$.\footnote{In fact any power of $y$, including none at all, is sufficient to form an equivalent Hamiltonian with differing lapse. In this construction we have explicitly chosen to keep the same time parametrization.} The dynamical system described by this Hamiltonian is exactly that of the contact system, but we recover conservation of the Hamiltonian and Liouville's theorem as the extra variable, $y$ acts to compensate for the non-conservative nature of the contact system. It is a straightforward exercise to show that Hamilton's equations for the variables $p,q$ and $S$ are the same as those we derive from the contact Hamiltonian, and that these form an autonomous system; although we have introduced $y$ as a coordinate we could evolve the system without ever making reference to it, simply by expressing the equations of motion for $p,q$ and $S$. Since we have constructed this system beginning with the equation of motion for $S$, it should be unsurprising to note that this does indeed return the Herglotz-Lagrangian, $\L^H$. We can finally take this Hamiltonian and perform a Legendre transform and thus obtain a Lagrangian on the tangent bundle over the extended coordinate space.

In the damped harmonic oscillator this construction is performed by taking $\pi = yp$ and forming the symplectic structure $\omega=dS\wedge dy + d\pi \wedge dx$. The Hamiltonian for this system is then:
\be \H= y\H^c = \frac{\pi^2}{2my} + \frac{ykx^2}{2} + \frac{y\mu S}{m} \ee
and thus we find the equations of motion from Hamilton's equations:
\be
\dot{x}=\frac{\pi}{my} \quad \dot{\pi}=-kyx \quad \dot{y}=\frac{\mu y}{m} \quad \dot{S}=\frac{\pi^2}{2my^2} - \frac{kx^2}{2} - \frac{\mu S}{m} 
\ee
Which we can express in terms of $p,q$ and $S$;
\be \dot{x} = \frac{p}{m} \quad \dot{p}=-kx-\frac{\mu p}{m} \quad \dot{S} = \frac{p^2}{2m} -\frac{kx^2}{2} - \frac{\mu S}{m} \ee
Thus as advertised $x,p$ and $S$ form an autonomous system. Further, since our system is linear in $S$ this can be evolved without reference to $S$, and reproduce $m\ddot{x}+\mu\dot{x}+kx=0$.  We also obtain $y= y_0 e^{\frac{\mu t}{m}}$ where $y_0$ is a constant. This final equation taken together with equation (\ref{DampedHamiltonianEvolution}) demonstrate the conservation of the Hamiltonian, $H$. Further we note that within this construction, as stated above, the equation of motion for $S$ is exactly that of equation (\ref{DampedHarmonicHerglotzLagrangian}). 

Once we have found the Hamiltonian for our system, it is simply a matter of performing the Legendre transform again to recover a Lagrangian on the tangent bundle over the extended configuration manifold. From the symplectification of the contact Hamiltonian and the definitions of the new momenta as represented in symplectic potential derived from the contact form, one can show directly that such a Lagrangian is:
\be
\L = y \left(\L^H(q,\dot{q},S) + S \frac{\dot{y}}{y} \right) \label{FinalLagrangian} 
\ee
and a direct application of the Euler-Lagrange equations gives the same equations of motion. The new coordinate, $y$ has a simple equation of motion; we can see from the construction that 
\be \dot{y}=- y \frac{\partial \L^H}{\partial S} \label{DummyEvolution} \ee
and thus so long as this is invertible the Lagrangian given in equation (\ref{FinalLagrangian}) will reproduce all of our equations of motion. In this form it is simple to see that the Lagrangian is linearly proportional to $y$, and hence changing $y$ by a factor will rescale the action, but leave the equations of motion for the variables $q,\dot{q}$ and $S$ unchanged. This is a example of dynamical similarity \cite{DynSim,Bravetti:2020jev}. Dynamical similarities are symmetries of a theory that are not standard canonical transformations \cite{Bravetti:2020jev}, under which the equations of motion for an autonomous system of physical observables are retained despite the rescaling of some quantities such as the Lagrangian, Hamiltonian and symplectic structure. A mathematical framework for how one can identify such symmetries and use them to construct the autonomous systems was given in \cite{DynSim}, a mathematical description of their nature and the induced Noether type symmetries discussed in \cite{Bravetti:2020jev}, and a pedagogical explanation of their nature and implications for understanding physics is explored in \cite{NewPaper}. 

In our example life is a little more complicated since the Hamiltonian is linear in $S$, and hence the equation of motion for $y$ is linear. However this is easily resolved by noting that the equation of motion allow us to replace $y$ by its time evolution. As such for the damped harmonic oscillator we see that the Lagrangian can be rendered in time dependent form following the Legendre transform, and is:
\be \L = e^{\frac{\mu t}{m}} \left(\frac{m\dot{x}^2}{2} - \frac{kx^2}{2} \right) \ee
The Euler-Lagrange equation for $x$ gives us back the initial damped harmonic oscillator.

In this section we have begun with an action principle following Herglotz, where the Lagrangian depends on the action as well as the tangent bundle over the configuration manifold. Through a Legendre transform we found the contact Hamiltonian, which we then symplectified to find a Hamiltonian representation of the same system in a larger space. This was then Legendre transformed again to give a Lagrangian with an interesting symmetry on an the tangent bundle over an extended configuration manifold. This entire process can be performed in reverse; we can begin with a Lagrangian which has a dynamical similarity and end up with a contact Hamiltonian and Herglotz action. In the following section we will show how this applies to the action of equation (\ref{FLRWHerglotzAction}) and gives us the Einstein-Hilbert action on flat FLRW cosmology, and how it can be extended to include curvature. In section \ref{Sec:AndBackAgain} we describe the inverse of this process.

\section{Equivalence with FLRW Cosmology}
\label{Sec:FLRW}

The Friedmann-Lema\^itre-Robertson-Walker (FLRW) cosmological models are homogeneous, isotropic solutions to Einstein's equations. The geometry of these models is encoded in the line element, 
\be
ds^2 = -dt^2 + a^2 \left(\frac{dr^2}{1-kr^2} + r^2 d\Omega^2 \right) \label{FLRWLineElement}
\ee
The time evolution of the geometry is thus encoded through a single variable, the scale factor $a$, and a constant the curvature $k$. It is well established that physics is insensitive to the value of the scale factor at any given time. It is typical to choose a specific event, such as the present time, to set the value of $a$ to one, and thus fix this ambiguity. However, this is simply a matter of convention; no physical observable is affected by this choice. Thus there is within the description of physics a redundancy in this choice. One of the key aims of this section will be to show how we can remove this redundancy and work with a smaller set of data which still reproduces the behaviour of all the observables. For simplicity, we will take here the matter component to be described by a Lagrangian density for a single field, $q$. This is trivial to generalize but a single field will serve to illustrate our point.

The Lagrangian for a FLRW cosmological system can be found by beginning with the Einstein-Hilbert action and restricting the set of space-times to those described by equation (\ref{FLRWLineElement}). In doing so, we will choose to work with the volume $v=a^3$ since this cleans up the algebra somewhat. Up to boundary terms which we will ignore as they do not play a role in bulk dynamics we find that the Lagrangian density is given
\be 
\L = \frac{-1}{24\pi G} \frac{\dot{v}^2}{v} + v \L_m (q,\dot{q}) - N v^{\frac{1}{3}}  \label{FLRWLagrangian}
\ee
Here $N$ is a constant which is zero in the case of a flat spatial slice, which is the first case we will consider. 

Let us now follow the procedure outlined in the previous section in the cosmological case. Our Herglotz-Lagrangian is given by equation (\ref{FLRWHerglotzAction}) and thus can be symplectified by introducing a new coordinate and extending the contact space to the tangent bundle over this extended coordinate space. In doing so we see that equation (\ref{DummyEvolution}) gives:
\be \frac{\dot{y}}{y} = - 3S \ee
and hence we identify $y$ with the volume $v$, up to a free choice of scale. We can find the Lagrangian that results from our symplectification, following equation (\ref{FinalLagrangian}) and see that it is indeed:
\be \L = y \L^H + S \dot{y} = v \left( - \frac{\dot{v}^2}{6v^2} + 4 \pi G \L_m \right) \ee
which is, up to an overall factor, that given in equation (\ref{FLRWLagrangian}). Thus we arrive at a key result: The Lagrangian derived from the Einstein-Hilbert action is equivalent to the symplectification of the Herglotz-Lagrangian given in equation (\ref{FLRWHerglotzAction}). This establishes that the observable content of both formulations are identical, yet the ontological setting of the Herglotz case is simpler, having one fewer degree of freedom in the description. 

In the case of non-flat spatial slices it would appear that there is an explicit dependence on $v$, and hence this would seem to require the reintroduction of the scale factor. However it turns out that this is not the case. Our dynamics exhibits friction, thus any kinetic term in the Herglotz Lagrangian this will decay even in the absence of a corresponding potential. In particular, adding an extra term to $\L^H$ proportional to $\dot{z}^{-2}$, where $\L^H$ is independent of $z$ reproduces the effect of curvature. This can be seen in the Euler-Lagrange equation for $\dot{z}$:
\be \frac{d}{dt} \left(\frac{\partial \L^H}{\partial \dot{z}} \right) - \frac{\partial \L^H}{\partial S} \frac{\partial \L}{\partial \dot{z}} =0 \rightarrow \dot{z} = \dot{z_0} e^{-\int S dt}  \ee
and hence $\dot{z}$ would have the same evolution as the scale factor. Thus we can include the effect of curvature by introducing a new kinetic only variable. This is generalizable to cover any apparent dependence on the scale factor by taking different powers of the kinetic term.

At this point it may seem that we have gained little by our exchanging of the scale factor for a new kinetic term; it would appear that we have removed a degree of freedom just to reintroduce it in a different form. However we should recall that in the standard picture we need to specify not just the value of the scale factor at a given time, but also the value of $k$, the curvature. Thus we have taken two pieces of data required to evolve our system (one initial condition, $a_0$, one constant, $k$) and replaced them by a single initial condition $z_0$. Further we should note that in the standard picture any constant such as $k$ can be replaced by a kinetic term with no conjugate position dependence in the Lagrangian, as this trivially gives rise to a constant evolution. Thus we can consider the role of initial conditions and constants interchangeable. In this manner we can replace any constant coupling to the unobservable scale factor with a kinetic term and initial condition for that term, both of which require the same amount of information to specify. 

\section{Extension to Bianchi Models}
\label{Sec:Bianchi}
The Bianchi models are a natural extension of the FLRW space-times considered thus far, as they relax the assumption of isotropy, allowing for different expansion rates along each of three spatial directions. The set of spatial slices (three dimensional homogeneous metrics) was mapped out by Bianchi. These models will serve a dual purpose for us. In the first instance they allow us to examine the evolution of geometric degrees of freedom other than just scale. In the second, it is believed that such models better capture the dynamics of general relativity in the neighbourhood of singularities \cite{BKL,AR,Uggla2,AHS}.

The Bianchi models are characterized by their Killing vector fields, $\xi_i$. There must exist three of these as our spatial manifold is three dimensional and homogeneous. We distinguish models by the Lie bracket of the $\xi_i$: 
\be [\xi_i,\xi_j] = C^k_{ij} \xi_k \ee
We further decompose the structure constants $C^k_{ij}$ through:
\be C^k_{ij} = \epsilon_{ijl}n^{kl} + \alpha_i \delta^k_j - \alpha_j \delta^k_i \ee
wherein $\epsilon$ is the alternating tensor and $\delta$ the Kronecker delta. $n^{ij}$ is a diagonal tensor which we can describe in terms of its eigenvalues $n^i$. When non-zero, these in turn can be set equal to $\pm 1$ without loss of generality through rescaling their lengths. The models for which $\alpha_i=0$ are called the class A models, and these are the models for which a Hamiltonian description exists, and those with which we will concern ourselves here.\footnote{The class B models, for which $\alpha_i \neq 0$, are not believed to be physically relevant in the vicinity of singularities and are of limited physical significance\cite{Ringstrom}.} Thus for the models we are interested in there exists a co-frame $\sigma^i$ compatible with the symmetries, for which 
\be d\sigma^1 = -n_1 \sigma^2 \wedge \sigma^3 \quad d\sigma^2 = -n_2 \sigma^3 \wedge \sigma^1 \quad d\sigma^3 = -n_3 \sigma^1 \wedge \sigma^2 \ee
The line element is parametrized in terms of this co-frame:
\be \label{BianchiLineElement} ds^2 = -dt^2 +a^2 \left(e^{-\phi_1 -\frac{2 \phi_2}{\sqrt{3}}} (\sigma^1)^2 + e^{-\phi_1 + \frac{2 \phi_2}{\sqrt{3}}} (\sigma^2)^2 + e^{2\phi_1} (\sigma^3)^2 \right) \ee
The Ricci scalar on the spatial manifold for such a metric can be decomposed into the product of a term dependent on the scale factor and a term that depends on the $\phi_i$ and the $n_i$:
\be {}^3R = \frac{V_s (\phi_1,\phi_2,\vec{n})}{a^2} \ee
wherein we shall call $V_s$ the shape potential. This is determined by the structure constants and the anisotropies, $\phi_1$ and $\phi_2$. Following Uggla \cite{Uggla3}, this is:
\be
V_s = \frac{1}{2} e^{4 \phi_1} h_-^2 + n_1 e^{-2 \phi_1} h_+ + \frac{1}{2} n_1^2 e^{-8 \phi_1}
\ee
with
\be
h_\pm = n_2 e^{2\sqrt{3} \phi_2} \pm n_3 e^{-2\sqrt{3} \phi_2}  
\ee
Hereafter we shall drop the $\vec{n}$ in the shape potential for simplicity, as this will be a constant on each space-time.
The Herglotz-Lagrangian for this system is then:
\be \label{BianchiHerglotzLagrangian} \L^H = \frac{3S^2}{2} +\frac{\dot{\phi}_1^2}{2}+\frac{\dot{\phi}_2^2}{2} - \frac{V_s^3(\phi_1,\phi_2)}{\dot{z}^2} +4\pi G \L_m \ee
There are several notable features of this. The first is that following the relaxation of isotropy, $\L^H$ remains closely related to that given in equation (\ref{FLRWHerglotzAction}). This is a not entirely surprising, as in the usual symplectic case the Lagrangians for the FLRW and Bianchi spacetimes are similar. However as with the coupling to matter, the interaction between the anisotropic parts and the mean extrinsic curvature is no enacted through the action itself, rather than through the minimal coupling. In other words, the reason that this system does not separate into two independent systems is because both contribute to the $S^2$ term, which in turn acts frictionally on both parts. The second thing to note is that the form of the shape potential $V_s$. This is the scalar curvature of the unit spatial three-manifold with anisotropies $\phi_1$ and $\phi_2$. In the Herglotz Lagrangian this appears in a term $V_s^3/\dot{z}^2$. On an initial, surface level, inspection one might suspect that this would give a distinct evolution from symplectic case. However this is not the case, as is verified by considering the equation of motion for $\dot{z}$. From the general form of the Herglotz equations (equation (\ref{HerglotzGeneralEOM})) we see:
\be \frac{d}{dt} \left(\frac{V_s^3}{\dot{z}^3}\right) = -3S \frac{V_s^3}{\dot{z}^3} \ee
which can be solved to find the dynamics of $\dot{z}$:
\be \dot{z} = \dot{z}_0 V_s e^{-\int S dt} \ee
and hence $V_s/\dot{z}$ takes on the same role as the scale factor would in the symplectic version. 

To show that equation (\ref{BianchiHerglotzLagrangian}) is equivalent to the Einstein-Hilbert action restricted to homogeneous space-times, we follow the prescription of equation (\ref{FinalLagrangian}). Again identifying the new coordinate introduced in symplectification with the volume, $v$ we find:
\be \label{HalfwayBianchi} \L = v\L^H + S\dot{v} = v\left( -\frac{\dot{v}^2}{6v^2} +\frac{\dot{\phi_1}^2}{2} +\frac{\dot{\phi_2}^2}{2} - \frac{V_s^3}{\dot{z}^2} +  4\pi G \L_m \right) \ee
Again this might initially appear different from the usual Lagrangian due to the presence of $\dot{z}$. However the Euler-Lagrange equation for $\dot{z}$ tells us that $\frac{vV_s^3}{\dot{z}^3}$ is a constant, hence $V_s/\dot{z}$ is proportional to $v^{1/3}$. Upon replacing the $\dot{z}$ term in equation (\ref{HalfwayBianchi}) using this, we recover the correct form.

\section{Revealing the contact system from the Einstein-Hilbert action}
\label{Sec:AndBackAgain}

In this paper we have presented the action in equation (\ref{FLRWHerglotzAction}) and shown that its symplectification is the Einstein-Hilbert action. This symplectification gives rise to a symmetry of the system under changing of the scale factor - this is a dynamical similarity between solutions. Here we will show how this action can be found beginning with the Einstein Hilbert action. The general procedure for this was laid out in \cite{DynSim}, and here we will apply it in the specific contexts of interest - the FLRW and Bianchi cosmologies. This is a three step process: First we identify the dynamical similarity either in the Lagrangian of Hamiltonian formulation. Then we use this to provide the contact system by expressing the Hamiltonian in terms of invariants of these transformations and thus constructing a contact Hamiltonian and contact form from these invariants. Finally we perform a Legendre transform to return the Herglotz Lagrangian.

We begin with the Einstein-Hilbert Lagrangian, $\L$, given in equation (\ref{FLRWLagrangian}). It is apparent from the form of $\L$ that there is a symmetry of this Lagrangian under which $\L \rightarrow \lambda \L$. Under the transformation $D$ which is given:
\be D:\{v,N\}\rightarrow \{\lambda v, \lambda^\frac{2}{3} N\} \quad \quad  D: \L \rightarrow \lambda \L  \ee
we see that the physical observables ($H=\frac{\dot{v}}{3v},q, \dot{q}$) are unchanged and the Lagrangian retains its form but is multiplied by an overall factor. The reason for this is related to the freedom to set the value of the scale factor at an event. Since different choices of the scale factor do not lead to distinct cosmological solutions, this transformation cannot affect the physical observables. The equations of motion come from extremizing the action. Since $D$ has not changed any of the physical observables their evolution remains unchanged under this transformation. Hence $D$ describes a map between solutions which are not physically distinguishable, and spans a one dimensional curve through through the space of $v,\dot{v},q,\dot{q},N$. Any observer within the cosmology could not locate their position on this curve. We could also describe $D$ as determining an equivalence class of cosmological solutions ${\gamma_i}$ with equivalence relation $\sim$ under which $\gamma_1 \sim \gamma_2$ if all observables of $\gamma_1$ and $\gamma_2$ agree. The invariants of $D$ are simple to identify - they are the matter variables $q,\dot{q}$, the Hubble parameter, $H$, and a combination of $N$ and $v$, $X=\frac{N^{\frac{3}{2}}}{v}$. 

Let us now construct the Hamiltonian for our theory. In this we note that the matter Lagrangian, $\L_m$ is multiplied by $v$ throughout, hence for our Legendre transform the conjugate momentum to $q$ will be 
\be \Pi =\frac{\partial \L}{\partial \dot{q}} = v \frac{\partial \L_m}{\partial \dot{q}} = vp \ee
since the matter Lagrangian is minimally coupled and hence independent of $v$, and wherein we denote by $p$ the conjugate momentum to $q$ of the uncoupled Lagrangian. If we denote by $\H_m$ the Hamiltonian obtained from the non-coupled Lagrangian, $\L_m$, alone it is simple to show that the Hamiltonian is
\be \H = v \left(-6 \pi P_v^2 + \H_m(q,\frac{\Pi}{v}) +Nv^{-\frac{2}{3}} \right) \label{FLRWHamiltonian}\ee
where $P_v=-\frac{H}{4\pi}$. The symplectic structure is $\omega = dP_v \wedge dv + d\Pi \wedge dq$. We can then express $D$ as a vector field $\D$ on phase space:
\be 
\D  = v\pv{v} + \Pi \pv{\Pi} +\frac{2}{3} N \pv{N}
\ee
The action of $\D$ is to change both the Hamiltonian and symplectic structure in a way that preserves their forms and hence does not affect the equations of motion - it is a non-strictly canonical transformation \cite{Carinena:2013zpa,Carinena:2014bda}:
\be \Lie_\D \H = \H \quad \Lie_\D \omega = \omega \ee
Hence again we see that points on phase space connected by integral curves of $\D$ represent indistinguishable cosmological solutions:
\be \Lie_\D H = \Lie_\D q = \Lie_\D p = \Lie_\D X = 0 \ee

We are now almost in position to find the contact Hamiltonian following \cite{DynSim}. However we note that one of our invariants, $X$ is composed of a product of a constant and a dynamical variable, and as such will have a time evolution. In \cite{DynSim} we have treated $\D$ as mapping between different points on phase space, but in this case it also changes the values of constants. To alleviate this, let us promote $X=N^{\frac{3}{2}}$ to be a momentum with no conjugate position. As such, Hamilton's equations tell us that it will be a constant in time, and thus this is an equivalent formulation of the problem. We thus extend the symplectic structure to include $dX \wedge dz$ in which $z$ is a dummy configuration variable that has no physical meaning. Then we can form the contact Hamiltonian through
\be \H^c = \frac{\H}{v} = -6 \pi P_v^2 + \H_m (q,p) + X^{\frac{2}{3}} \ee
with contact form 
\be \eta = \frac{\iota_\D \omega}{v} = -dP_v + p dq + X dz \ee

From these we can find the equations of motion for our variables. These differ from Hamilton's equations as the system involves friction - the general form of the equations of motion for a contact Hamiltonian with contact form $-dA + \sum_i y_i dx_i$ are:
\be  \dot{x}^i = \frac{\partial \H^c}{\partial y_i} \quad \dot{y_i} = -\frac{\partial \H^c}{\partial x_i} - y^i \frac{\partial \H^c}{\partial A} \quad \dot{A}= y_i \frac{\partial \H^c}{\partial y_i} - \H^c \ee
and hence we find our equations of motion are:
\ba \dot{q} &=& \frac{\partial \H_m}{\partial p} \quad \dot{p} = -\frac{\partial \H_m}{\partial q} + 12 \pi P_v p \quad  \dot{X}= 12\pi P_v X \\
 \dot{P_v} &=& p\frac{\partial \H_m}{\partial p} - \H_m + 6\pi P_v^2 - \frac{X^\frac{2}{3}}{3} \quad \dot{z}=\frac{2}{3X^{\frac{1}{3}}} \ea
Within this we see that the equation of motion for $P_v$ contains a term that is the Legendre transform of the uncoupled matter Lagrangian $\L_m$. Further, as we note the equation of motion for $P_v$ is the Legendre transform of the coupled system - it is precisely the Herglotz Lagrangian $\L^H$ given in equation (\ref{FLRWHerglotzAction}), with the curvature term added. We can thus rewrite this in terms of the velocities $\dot{q}$ and $\dot{z}$ to recover:
\be \dot{P_v} = 6\pi P_v^2 + \L_m -\frac{3}{4 \dot{z}^2} \ee
Which is the Herglotz Lagrangian of equation (\ref{FLRWHerglotzAction}) if we identify $S=4 \pi P_v$, i.e. $S=-H$. 

\section{Discussion}
\label{Sec:Discussion}

We have shown that a complete description of cosmological systems can be obtained from an action principle that makes no reference to the scale factor. Thus we can construct all the observables of cosmology, in the usual FLRW and Bianchi models, without ever making reference to size. Whilst it is true that one can always embed such systems in models in which scale is made explicit, this is never strictly necessary in finding the evolution of physical observables. The action principle employed is of the type discussed by Herglotz, and thus describes a contact system which is inherently a frictional system. We have shown that the symplectification of this action reproduces exactly the Einstein-Hilbert action for cosmological systems, and thus we reproduce exactly the dynamics of relativistic cosmology and all physical observables thereon.

It is interesting to note that the cosmological dynamics of a matter system can be obtained by beginning with the matter Hamiltonian (or Lagrangian) and ``contactifying" it. Here we closely follow a construction given by Arnold \cite{ArnoldBook}. Consider a the cotangent bundle over configuration space, $T^*M$. We can construct the ``contactification" by taking a bundle with fibres $\mathbb{R}$ over the base space of $T^*M$. This is isomorphic to the space $\mathcal{C}$ on which the FLRW dynamics takes place. In local coordinates we can express this is a direct product $\mathbb{R}\times T^*M$ with coordinates $S,p,q$. If the symplectic form is exact (i.e. there exists a symplectic potential $\theta$) then we can give a canonical contact form $\alpha=\theta-dS$. Then we recover cosmological dynamics by taking $\H^c = \H_m - \kappa S^2$. We can then re-symplectify this system in exactly the way described in section \ref{Sec:FLRW} - introducing a new coordinate to parametrize the space of contact form. In other words, if we take a matter system and introduce a quadratic friction term by first extending the phase space by taking its product with the reals, then adding an extra term proportional to a coordinate on this fibre squared, together with taking a contact form from the symplectic potential and an exact form on this fibre, we reproduce the dynamics of an expanding universe. Thus the usual Einstein-Hilbert action and the dynamics of FLRW geometry are introduced by taking a matter system, contactifying it to introduce friction, then re-symplectifying the result. 

On one hand it should be somewhat unsurprising that we can formulate cosmology without ever referencing scale. After all, the scale factor is known to play no direct role in physics, and it has long been understood that there are multiple representations of the same physical system that differ only by the choice of scale factor at a given event (a choice that can be made exactly once per representation). Since in many cases this is measured with reference to a fiducial cell, the size of which is also an arbitrary choice, it is clear that physics should have no dependence upon this choice. On the other hand, general relativity, from which our cosmological systems are derived, is the dynamics of geometry. It would seem apparent that this geometry should contain a notion of scale. If our cosmological conclusions carry over to an action for general relativity in full generality this would indicate quite strongly the distinction between size, as measured through the determinant of a metric on a spatial manifold, say, and shape measured in terms of relative shear. It is not a priori obvious that this should be the case; the homogeneity that we impose in cosmological models forces there to be no distinction between points on a spatial manifold and therefore no two distinguishable points between which a notion of distance could be established. When this assumption is relaxed, it is clear that such a notion can be well defined, for example in terms of the distance between the ends of the metre des archives in Paris. Nonetheless, there exists a transformation rescaling all such distances between points (together with definitions of a second and values of the fundamental constants) which preserves the form of the Einstein-Hilbert action, therefore it seems likely such a reduced system will exist.

The removal of scale from our ontology has several important consequences. It has allowed us to construct a more parsimonious theory, requiring one fewer initial datum to provide a complete evolution. This follows the recently coined `principle of essential and sufficient autonomy' (PESA) which posits that when considering two (or more) viable models, if all else is equal then the one which requires fewer external inputs is to be preferred \cite{NewPaper}. To quote Ismael and van Fraassen \cite{Ismael}, ``Formalisms with little superfluous structure are nice, of course, because they reflect cleanly the structure of what they represent; they have fewer extra mathematical hooks on which to hang the mental structures that we project onto the phenomena."  The new action, being independent of the scale factor has eliminated one such hook. 

The FLRW Lagrangian, shown in equation (\ref{FLRWLagrangian}), involves the fields $v$ and $q$ and their time derivatives, and the standard Euler-Lagrange equations give rise to the usual dynamics. These are given:
\ba \frac{d}{dt} \left(\frac{\partial \L_m}{\partial \dot{q}} \right) + \frac{\dot{v}}{v} \frac{\partial \L_m}{\partial \dot{q}} - \frac{\partial \L_m}{\partial q} &=& 0 \\
\frac{1}{12 \pi G} \left(\frac{\ddot{v}}{v} + \frac{\dot{v}^2}{v^2} \right) + \L_m - \frac{N}{3v^{\frac{2}{3}}} &=& 0 \ea

At this point it would appear that we require five pieces of information to reproduce a solution to the system - the fields, their time derivatives and the value of $N$. This system is subject to a constraint - the Hamiltonian must vanish. Thus we should expect we need to specify four quantities to uniquely determine a physical solution. However we have shown that if we restrict our interests to be only the evolution of physical observables it will suffice to set only three. 

The notions of typicality or probability in cosmology are reliant upon the formulation of a measure by which the number of distinct space-times can be counted. As we have argued in the past \cite{Corichi:2013kua}, space-times which are distinguished only by the choice of scale factor at a given time should not be considered separate entities, and thus measures that involve either directly or indirectly an integral over scale factors are multiply counting solutions. In the standard framework of general relativity, this leads to an ambiguity as this direction is unbounded and cut-offs must be employed to render any counting finite. However the imposition of such cut-offs does not commute with the time-evolution of the system, hence measures focus or unfocus as the universe expands.\footnote{This is a somewhat subtle issue and often overlooked as Liouville's theorem is invoked without considering the evolution of cut-offs.} A measure on the contact space, however, inherits its finiteness (or lack thereof) in large part from that of the matter component. The Liouville measure on contact space \cite{Bravetti2} when pulled back to a surface of constant Hubble becomes the Liouville measure on the matter component evaluated at a given energy. This in turn avoids the infinities associated with integrals over the scale factor. 

The role of friction in measure focussing provides new perspectives on the problems of the initial state of the universe \cite{Gryb:2020wat}. Friction is by its very nature a time asymmetric process. At a generic point in the evolution of a system there is a direction of time in which the mechanical energy is increasing, and one in which it decreases. Similarly there is a direction of time in which measure are focusing and one in which they are not. Thus from the frictional behaviour of our system we can infer an arrow of time. Further, since measures are not preserved under time evolution the problem of the low entropy state of the early universe is alleviated

As we have recently proven \cite{Through,Mercati:2019cbn,Sloan:2019wrz}, the mathematical barrier to extending classical cosmological solutions beyond the initial singularity lies in the failure of the system of equations to be Lifschitz continuous, and hence satisfy the conditions of the Picard-Lindel\"off theorem. This problem is alleviated when scale is removed from the system; it is only the evolution of the scale factor that is ill-defined. Hence a scale free system has a unique, deterministic classical evolution through the big bang. Thus reports of general relativity predicting its own demise may be greatly exaggerated. This has been established through considering the relational evolution of Bianchi systems \cite{Through}, including those with inflationary matter \cite{Mercati:2019cbn} and FLRW cosmologies with scalar fields \cite{Sloan:2019wrz}. The initial singularity may be thought of as a point at which dynamical evolution reaches the boundary of the description of space-time in terms of a four dimensional Riemannian manifold. From the perspective of expansion as friction we gain a different intuition; it is the point at which an infinite amount of mechanical energy has been added to the system. Whilst this would appear to be also problematic as it again invokes infinities, when rendered in relational terms this can be considered as potential terms being subdominant to kinetic terms, and thus motion becoming geodesic on a relational space. 

The evolution of the scale factor plays a central role in quantum approaches to cosmology. Both the Wheeler-DeWitt quantization and Loop Quantum Cosmology take as one of their physical observables a term representing the volume of the universe, as measured against a fiducial cell. As our action principle has no such terms within it, this suggests that an alternative quantization may be performed. The action of equation (\ref{FLRWHerglotzAction}) is frictional, and therefore may be amenable to alternative quantizations. It is of particular interest to examine the quantization of frictional systems following, for example, the Lindblad equation.

\section*{Acknowledgements}

The author is grateful to Julian Barbour, Sean Gryb and Flavio Mercati for helpful discussions and comments.

\bibliographystyle{ieeetr}
\bibliography{HerglotzCosmology}

\begin{thebibliography}{10}

\bibitem{PII}
P.~Forrest, ``{The Identity of Indiscernibles},'' in {\em The {Stanford}
  Encyclopedia of Philosophy} (E.~N. Zalta, ed.), Metaphysics Research Lab,
  Stanford University, winter 2020~ed., 2020.

\bibitem{weyl2009philosophy}
H.~Weyl, O.~Helmer, and F.~Wilczek, {\em Philosophy of Mathematics and Natural
  Science}.
\newblock Princeton University Press, 2009.

\bibitem{Gibbons:1986xk}
G.~Gibbons, S.~Hawking, and J.~Stewart, ``{A Natural Measure on the Set of All
  Universes},'' {\em Nucl.Phys.}, vol.~B281, p.~736, 1987.

\bibitem{Measure}
A.~Ashtekar and D.~Sloan, ``{Loop quantum cosmology and slow roll inflation},''
  {\em Phys.Lett.}, vol.~B694, pp.~108--112, 2010.

\bibitem{Corichi:2010zp}
A.~Corichi and A.~Karami, ``{On the measure problem in slow roll inflation and
  loop quantum cosmology},'' {\em Phys.Rev.}, vol.~D83, p.~104006, 2011.

\bibitem{Measure2}
A.~Ashtekar and D.~Sloan, ``Probability of inflation in loop quantum
  cosmology,'' {\em General Relativity and Gravitation}, vol.~43, no.~12, 2011.

\bibitem{Corichi:2013kua}
A.~Corichi and D.~Sloan, ``{Inflationary Attractors and their Measures},'' {\em
  Class.Quant.Grav.}, vol.~31, p.~062001, 2014.

\bibitem{Barbour2014}
J.~Barbour, T.~Koslowski, and F.~Mercati, ``Identification of a gravitational
  arrow of time,'' {\em Phys. Rev. Lett.}, vol.~113, p.~181101.

\bibitem{Barbour2015}
J.~Barbour, T.~Koslowski, and F.~Mercati, ``Entropy and the typicality of
  universes,''

\bibitem{FlavioSDbook}
F.~Mercati, {\em Shape Dynamics: Relativity and Relationalism}.
\newblock Oxford Univ. Press, 2018.
\newblock (early version on the arXiv).

\bibitem{Through}
T.~A. Koslowski, F.~Mercati, and D.~Sloan, ``{Through the big bang: Continuing
  Einstein's equations beyond a cosmological singularity},'' {\em Phys. Lett.},
  vol.~B778, pp.~339--343, 2018.

\bibitem{geiges2008introduction}
H.~Geiges, {\em An introduction to contact topology}, vol.~109.
\newblock Cambridge University Press, 2008.

\bibitem{ContactIntro}
J.~B. {Etnyre}, ``{Introductory Lectures on Contact Geometry},'' Nov. 2001.

\bibitem{Bravetti}
A.~{Bravetti}, H.~{Cruz}, and D.~{Tapias}, ``{Contact Hamiltonian mechanics},''
  {\em Annals of Physics}, vol.~376, pp.~17--39, Jan 2017.

\bibitem{Bravetti2}
A.~{Bravetti} and D.~{Tapias}, ``{Liouville{\textquoteright}s theorem and the
  canonical measure for nonconservative systems from contact geometry},'' {\em
  Journal of Physics A Mathematical General}, vol.~48, p.~245001, Jun 2015.

\bibitem{Leon}
M.~{Lainz Valc{\'a}zar} and M.~{de Le{\'o}n}, ``{Contact Hamiltonian
  Systems},'' {\em arXiv e-prints}, p.~arXiv:1811.03367, Nov 2018.

\bibitem{ArnoldBook}
V.~I. {Arnold} and S.~P. {Novikov}, {\em {Dynamical systems IV. Symplectic
  geometry and its applications}}.
\newblock VINITI, 2001.

\bibitem{DynSim}
D.~Sloan, ``{Dynamical Similarity},'' {\em Phys. Rev.}, vol.~D97, no.~12,
  p.~123541, 2018.

\bibitem{Bravetti:2020jev}
A.~Bravetti and A.~Garcia-Chung, ``{A geometric approach to the generalized
  Noether theorem},'' 9 2020.

\bibitem{NewPaper}
S.~Gryb and D.~Sloan, ``{Similarity, indiscernibility, and the implications for
  physics},'' {\em Forthcoming}, 2020.

\bibitem{BKL}
V.~a. Belinsky, I.~m. Khalatnikov, and E.~m. Lifshitz, ``{A General Solution of
  the Einstein Equations with a Time Singularity},'' {\em Adv. Phys.}, vol.~31,
  pp.~639--667, 1982.

\bibitem{AR}
L.~Andersson and A.~D. Rendall, ``{Quiescent cosmological singularities},''
  {\em Commun. Math. Phys.}, vol.~218, pp.~479--511, 2001.

\bibitem{Uggla2}
C.~Uggla, ``{The Nature of generic cosmological singularities},'' in {\em
  {Recent developments in theoretical and experimental general relativity,
  gravitation and relativistic field theories. Proceedings, 11th Marcel
  Grossmann Meeting, MG11, Berlin, Germany, July 23-29, 2006. Pt. A-C}},
  pp.~73--89, 2007.

\bibitem{AHS}
A.~Ashtekar, A.~Henderson, and D.~Sloan, ``{A Hamiltonian Formulation of the
  BKL Conjecture},'' {\em Phys. Rev.}, vol.~D83, p.~084024, 2011.

\bibitem{Ringstrom}
H.~Ringstrom, ``{The Bianchi IX attractor},'' {\em Annales Henri Poincare},
  vol.~2, pp.~405--500, 2001.

\bibitem{Uggla3}
C.~Uggla, ``{Hamiltonian Cosmology},'' in {\em {Dynamical Systems in
  Cosmology}}, pp.~212--228, 1997.

\bibitem{Carinena:2013zpa}
J.~F. Cari\~nena, F.~Falceto, and M.~F. Ra\~nada, ``{Canonoid transformations
  and master symmetries},'' 2013.

\bibitem{Carinena:2014bda}
J.~F. Cari\~nena, I.~Gheorghiu, E.~Mart\'inez, and P.~Santos, ``{Conformal
  Killing vector fields and a virial theorem},'' {\em J. Phys.}, vol.~A47,
  no.~46, p.~465206, 2014.

\bibitem{Ismael}
J.~Ismael and B.~van Fraasen, ``{Symmetry as a guide to superfluous theoretical
  structure},'' in {\em {Symmetries in Physics: Philosophical Reflections}},
  pp.~371--392, 2002.

\bibitem{Gryb:2020wat}
S.~Gryb, ``{New Difficulties for the Past Hypothesis},'' 6 2020.

\bibitem{Mercati:2019cbn}
F.~Mercati, ``{Through the Big Bang in inflationary cosmology},'' {\em JCAP},
  vol.~10, p.~025, 2019.

\bibitem{Sloan:2019wrz}
D.~Sloan, ``{Scalar Fields and the FLRW Singularity},'' {\em Class. Quant.
  Grav.}, vol.~36, no.~23, p.~235004, 2019.

\end{thebibliography}

\end{document}